\documentstyle[newarcrc,fleqn]{article}
\input{psfig.sty}
\def\msun{{\rm M_{\odot}}}
\def\be{\begin{equation}}
\def\ee{\end{equation}}
\def\mdot{\dot M}

\def\msunyr{{\rm M_\odot yr^{-1}}}
\def\pcrit{P_{\rm crit}}
\def\mconv{M_{\rm conv}}
\def\ga{\,\hbox{\hbox{$ > $}\kern -0.8em \lower 1.0ex\hbox{$\sim$}}\,}
\def\la{\,\hbox{\hbox{$ < $}\kern -0.8em \lower 1.0ex\hbox{$\sim$}}\,}

\title{Secondary Stars in CVs: The Theoretical Perspective}

\author{Ulrich Kolb\address{Dept.\ of Physics \& Astronomy, University
        of Leicester, Leicester LE1 7RH, UK}
        \thanks{Present address: Dept.\ of Physics, The Open
        University, Walton Hall, Milton Keynes MK7 6AA} and
        Isabelle Baraffe\address{Ecole Normale Sup\'{e}rieure de Lyon,
        \\C.R.A.L.\ (UMR 5574 CNRS), F-69364 Lyon Cedex 07, France}}

\begin{document}
\maketitle

\begin{abstract}
We apply the new generation of theoretical models of low--mass stars
to secondaries in CVs, focussing on systems above the period gap. The
models confirm that the spectral type should be a good indicator of
the donor mass. The orbital period--spectral type diagram can
potentially constrain the long--term mean mass transfer
rate. A transfer rate that increases with decreasing period is most
easily reconciled with the observational data.
\end{abstract}

\vspace*{-80mm}

\noindent {\sf Invited review to appear in the proceedings of the
Warner Symposium on Cataclysmic Variables, \\ {\it New Astronomy
Reviews}, 1999, eds.\ P.A.~Charles, A.R.~King \&\ D.~O'Donoghue}
\vspace*{75mm}

\section{Introduction}

The new generation of low--mass star and brown dwarf models by Baraffe
et al.\ (1995, 1997, 1998, henceforth summarized as BCAH) represent 
a significant improvement in the quantitative description of 
stars with mass $\la 1$ $\msun$.

The main strengths of the models are in two areas: the microphysics 
determining the equation of state (EOS) in the stellar interior, and
the realistic non--grey atmosphere models which enter as the outer boundary
condition. The EOS (Saumon, Chabrier and Van Horn 1995) is
specifically calculated for very low--mass stars, brown dwarfs and
giant planets. It has been successfully tested against
recent laser--driven shock wave experiments performed at
Livermore, which probe the complex regime of pressure dissociation
and ionization relevant for these objects (cf.\ Saumon et al.\ 1998).  
Recent much improved cool atmosphere models (see e.g.\ the review of
Allard et al.\ 1997) now provide realistic atmosphere profiles, 
which we use as the outer boundary condition,
and synthetic spectra. Chabrier \& Baraffe (1997) have shown that
evolutionary models employing a grey 
atmosphere instead, e.g.\ the standard Eddington approximation,
overestimate the effective temperature for a given mass, and
yield too large a minimum hydrogen burning mass. 

Several observational tests confirm the success of evolutionary models
based on these improvements, e.g.\ mass--magnitude relations,
colour--magnitude diagrams (Baraffe et al. \ 1997, 1998), mass--spectral
type relations (Baraffe \& Chabrier 1996), the first cool BD
GL~229B (Allard et al.\ 1996), properties of components in detached
eclipsing binaries (Chabrier \& Baraffe 1995) and of field M--dwarfs
(cf.\ Beuermann et al.\ 1998). 

Here we consider the BCAH models in the context of CV secondaries, and
focus on the relation between spectral type ($SpT$), donor mass
($M_2$) and orbital period ($P$). We obtain the spectral type of a
stellar model from its calculated colour $(I-K)$ and the empirical $SpT 
- (I-K)$ relation established by Beuermann et al.\ (1998). A summary
of the input physics used for the calculations presented below is given
by Kolb \& Baraffe (1999).

\section{The mass--spectral type relation}

Mass loss causes 
stars to deviate from thermal equilibrium. The surface luminosity is
no longer in balance with the luminosity generated in the core by
nuclear reactions, and the difference causes the star to expand or
contract.   
Therefore the stellar radius can be either larger or smaller than 
the corresponding equilibrium radius (see e.g.\ Whyte \& Eggleton
1980). Remarkably, the effective temperature of low--mass
main--sequence stars is rather insensitive to the degree of thermal
disequilibrium. They behave
just like giant stars on the Hayashi line, expanding along an 
evolutionary track with nearly constant effective temperature.
This can be understood in terms of simple homology
relations for predominantly convective low--mass stars 
(e.g\ Stehle et al.\ 1996).

Using BCAH models we quantify the deviation from equilibrium spectral
type, as a measure of the surface temperature, for CV donors with
various evolutionary histories. We consider the following simple
cases: 
\begin{description}
\item{{\em Standard sequence:}} mass transfer starts from an unevolved
(ZAMS) donor with mass 1 $\msun$, proceeds at a constant rate
$1.5\times 10^{-9}$ $\msunyr$, stops when the donor
becomes fully convective (at mass 0.21 $\msun$), and resumes at the
lower rate $5\times 10^{-11}$ $\msunyr$ once the donor has settled
back into thermal equlibrium. This sequence fits the period gap in
the framework of disrupted orbital braking (e.g.\ King 1988,
Kolb 1996). 
\item{{\em Unevolved sequences with constant $\mdot$:}} 
mass transfer starts from an unevolved (ZAMS) donor with mass 1
$\msun$ and proceeds at a constant rate $\dot M$ ($10^{-8}$ $\msunyr$
or $10^{-7}$ $\msunyr$).  
\item{{\em Evolved sequences with constant $\mdot$:}} mass transfer
initiates from a donor which has already burned a significant fraction
of its hydrogen supply, but is still in the core hydrogen burning
phase. We show three examples: a moderately evolved low $\mdot$
sequence (initial central hydrogen content $X_c=0.16$, initial donor
mass $M_2=1.0$ $\msun$, $\mdot = 1.5 \times 10^{-9}$ $\msunyr$),
an evolved, low $\mdot$ sequence ($X_c=4\times10^{-4}$, $M_2=1.2$ $\msun$,
$\mdot = 1.5 \times 10^{-9}$ $\msunyr$), and an evolved high $\mdot$ 
sequence ($X_c=4\times10^{-4}$, $M_2=1.0$ $\msun$, $\mdot = 5 \times
10^{-8}$ $\msunyr$).   
\end{description}
Note that as the mass loss timescale is short compared with the nuclear
time, the nuclear state of the donor is essentially frozen once mass
transfer has started. 

\begin{figure}[tb]
\begin{minipage}[t]{77mm}
\psfig{figure=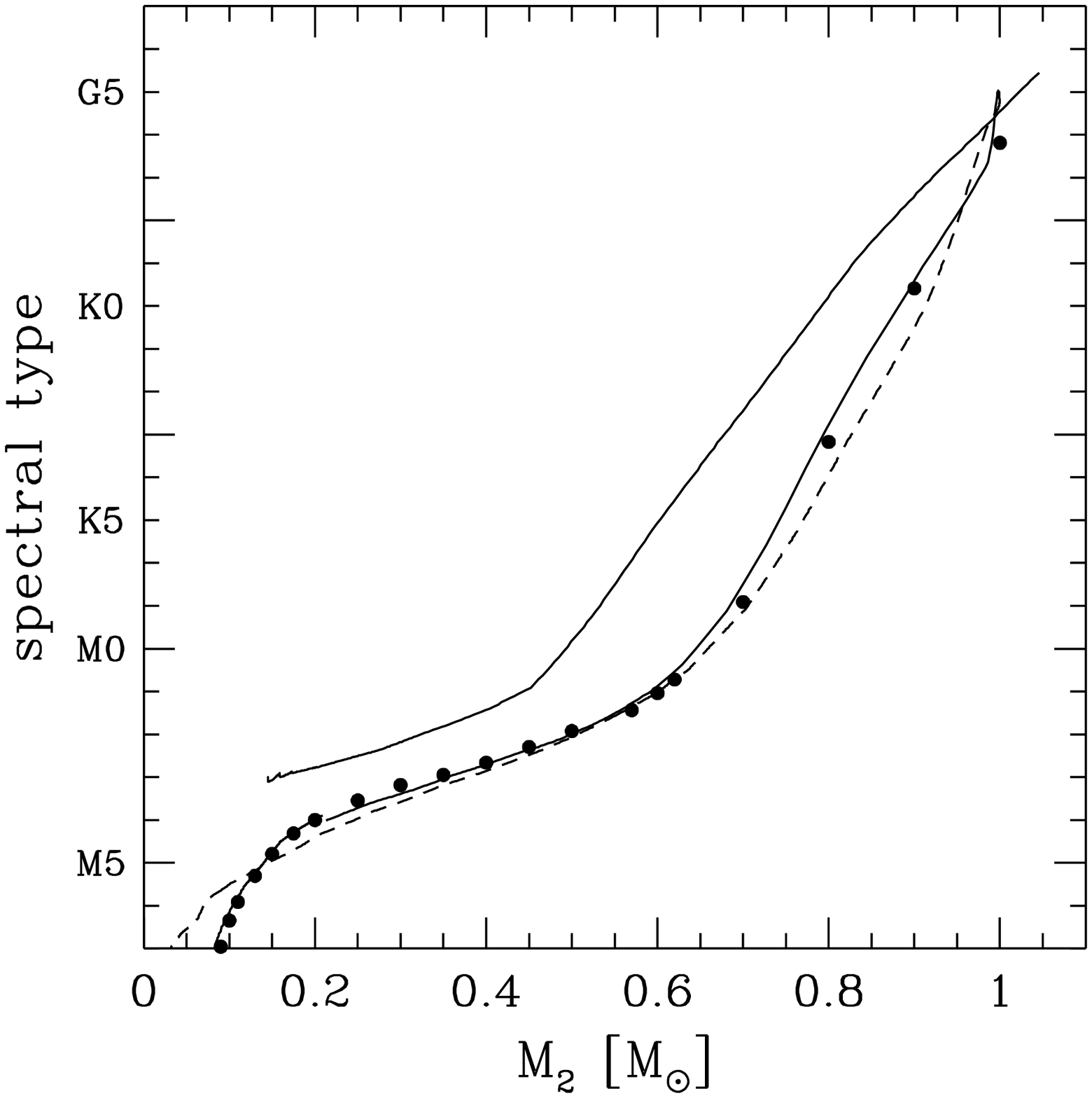,width=7.9cm,angle=0}
\vspace*{-0.5cm}
\caption{Spectral type versus secondary mass. Dots
denote ZAMS models. Solid curve, close to the dots: standard
sequence; solid curve, above the dots: donor moderately evolved
($X_c=0.16$). Dashed: donor unevolved, high $\mdot$ ($10^{-8}$
$\msunyr$).  
\label{fig:spm1}
}
\end{minipage}
\hspace{\fill}
\begin{minipage}[t]{77mm}
\psfig{figure=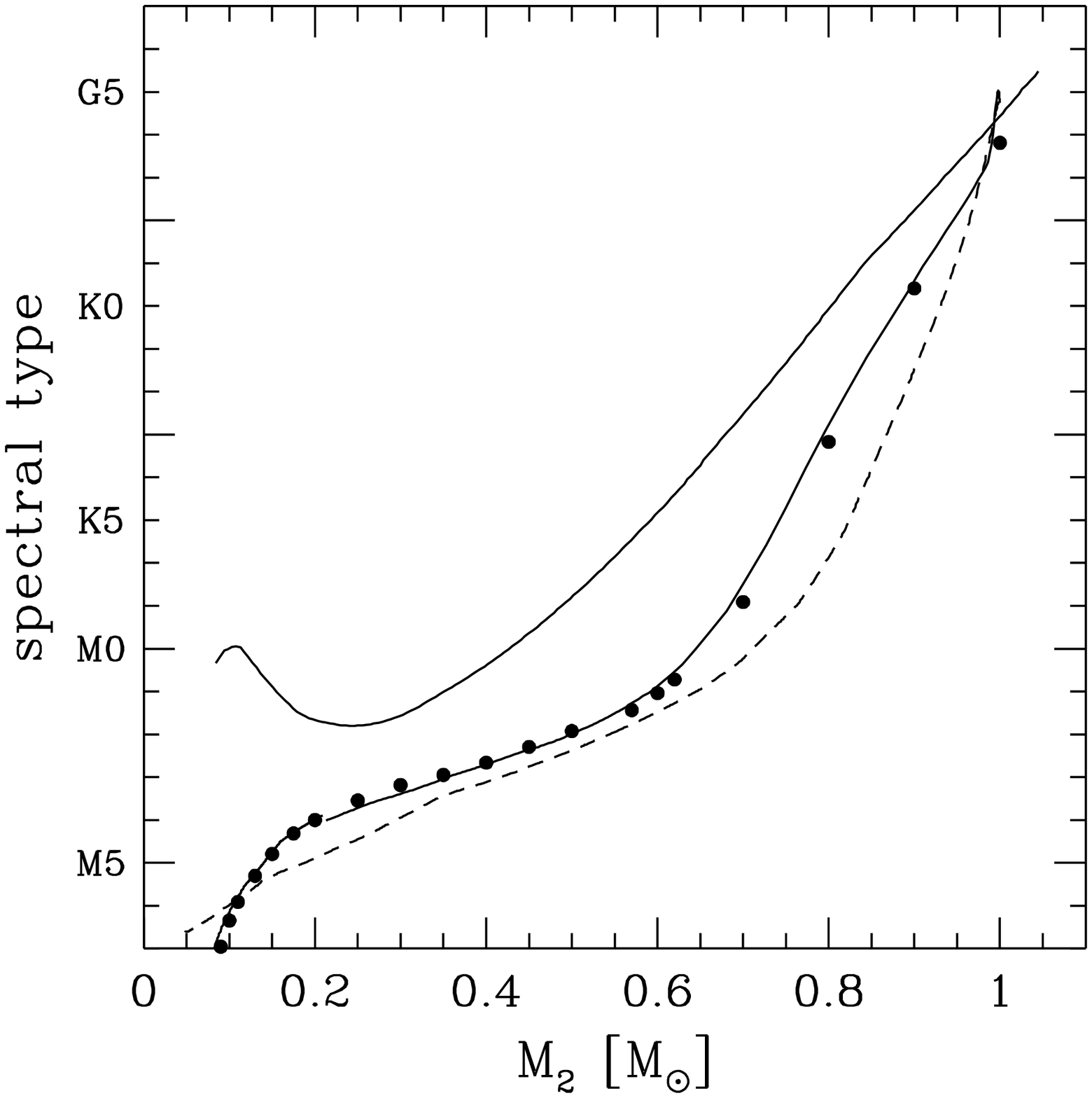,width=7.9cm,angle=0}
\vspace*{-0.5cm}
\caption{Same as Fig.~\protect{\ref{fig:spm1}}, but with extreme
assumptions for the evolved sequence (low $\dot M$,
$X_c=4\times10^{-4}$) and the unevolved high $\mdot$ sequence
($10^{-7}$ $\msunyr$).   
\label{fig:spm2}
}
\end{minipage}
\end{figure}

Fig.~\ref{fig:spm1} shows that the effect of thermal disequilibrium is
negligible along the standard sequence; this hardly
differs from that for CVs with a ZAMS donor. If $\mdot$ is higher than
in the 
standard sequence, the spectral type is slightly later for a given
mass, while in a sequence with evolved donors it is somewhat earlier. As
most CVs should form with essentially unevolved donors (Politano 1996,
de~Kool 1992), and as the 
estimated $\mdot$ exceeds $10^{-8}$ $\msunyr$ only for very few CVs
(e.g.\ Warner 1995, his Fig.~9.8), the upper solid curve ($X_c=0.16$)
and the dashed curve ($\mdot=10^{-8}$ $\msunyr$) in Fig.~\ref{fig:spm1}
should bracket the location of the majority of CVs in the
mass--spectral type diagram. The figure suggests that the spectral
type is a good indicator of the donor mass, with a 
typical uncertainty $\Delta M_2 \simeq 0.1\msun$, for any given
$SpT$ earlier than M0-M2. For later $SpT$ this mass ``determination''
becomes impractical as the curves in the figure  
flatten and $\Delta M_2$ is large. Even with extreme evolutionary
cases, the unevolved sequence with very high transfer rate
($\mdot=10^{-7}$ $\msunyr$) more typical for supersoft X--ray binaries
than for CVs, and the evolved sequence starting on the terminal
main--sequence, the range covered by the secondary stars in the 
$SpT$--mass diagram remains surprisingly narrow 
(Fig.~\ref{fig:spm2}; see Kolb, King \& Baraffe 1999 for more
details).  

Although the observations compiled by Smith \& Dhillon (1998) do
not seem to support this prediction, the observational errors on $M_2$
are far too large for any definitive conclusions to be drawn from this
mismatch.  
On the contrary, with improved and more reliable mass determinations,
one may hope to constrain the range of $\mdot$ and the relative
importance of evolved and unevolved systems in the CV population.

The figures also apply to secondaries in LMXBs. Evolutionary
considerations suggest that, unlike CVs, most short--period LMXBs,
form with an evolved main--sequence donor with mass $\ga 1$ $\msun$
(King \& Kolb 1997, Kalogera et al.\ 1998).

\section{The orbital period--spectral type diagram} 

\begin{figure}[htb]
\begin{minipage}[t]{77mm}
\psfig{figure=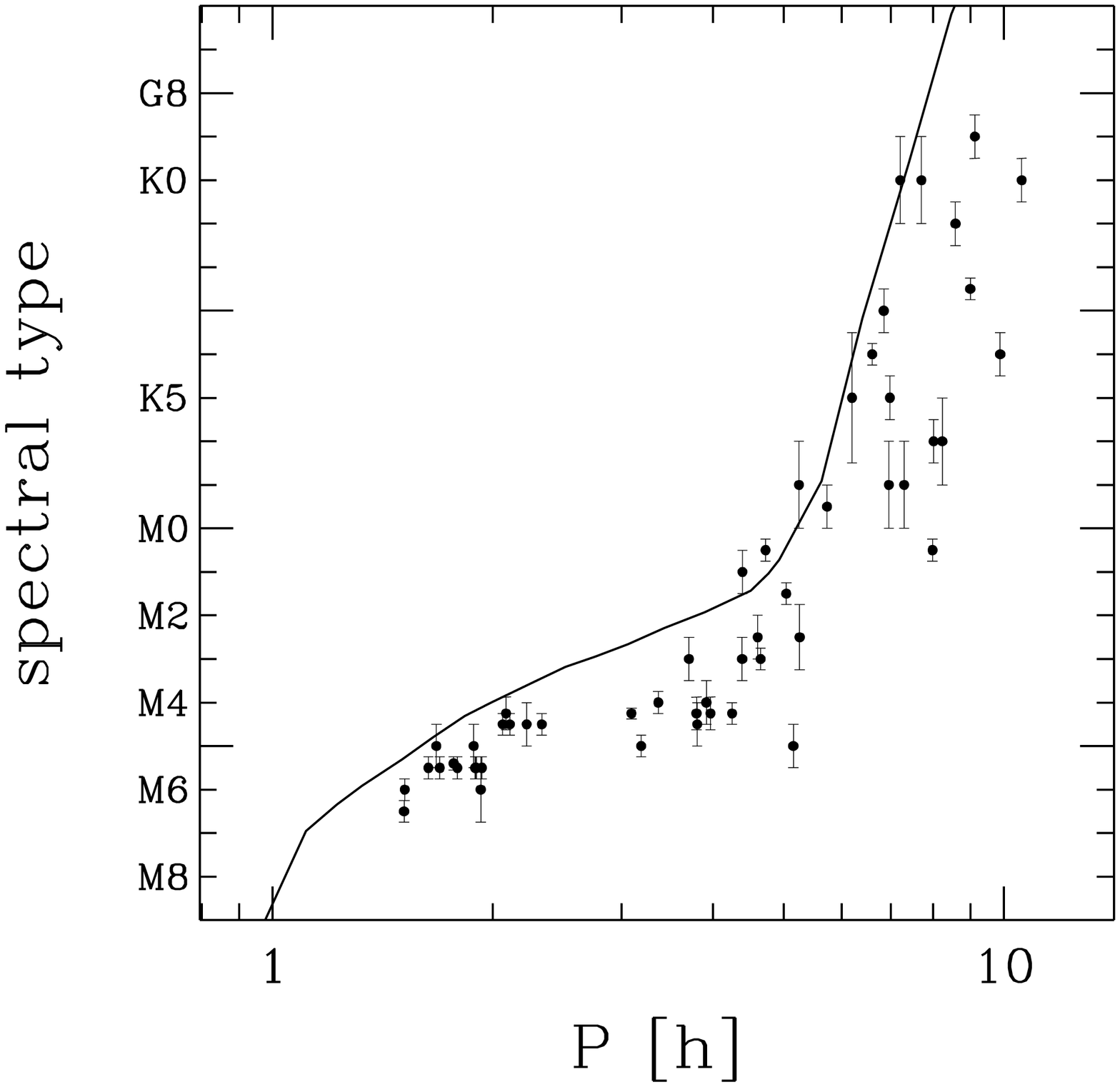,width=7.9cm,angle=0}
\vspace*{-0.5cm}
\caption{Spectral type versus orbital period for CV secondaries. Data
taken from Beuermann et al.\ 1998. Solid: theoretical location of 
ZAMS secondaries.
\label{fig:sppdata}
}
\end{minipage}
\hspace{\fill}
\begin{minipage}[t]{77mm}
\psfig{figure=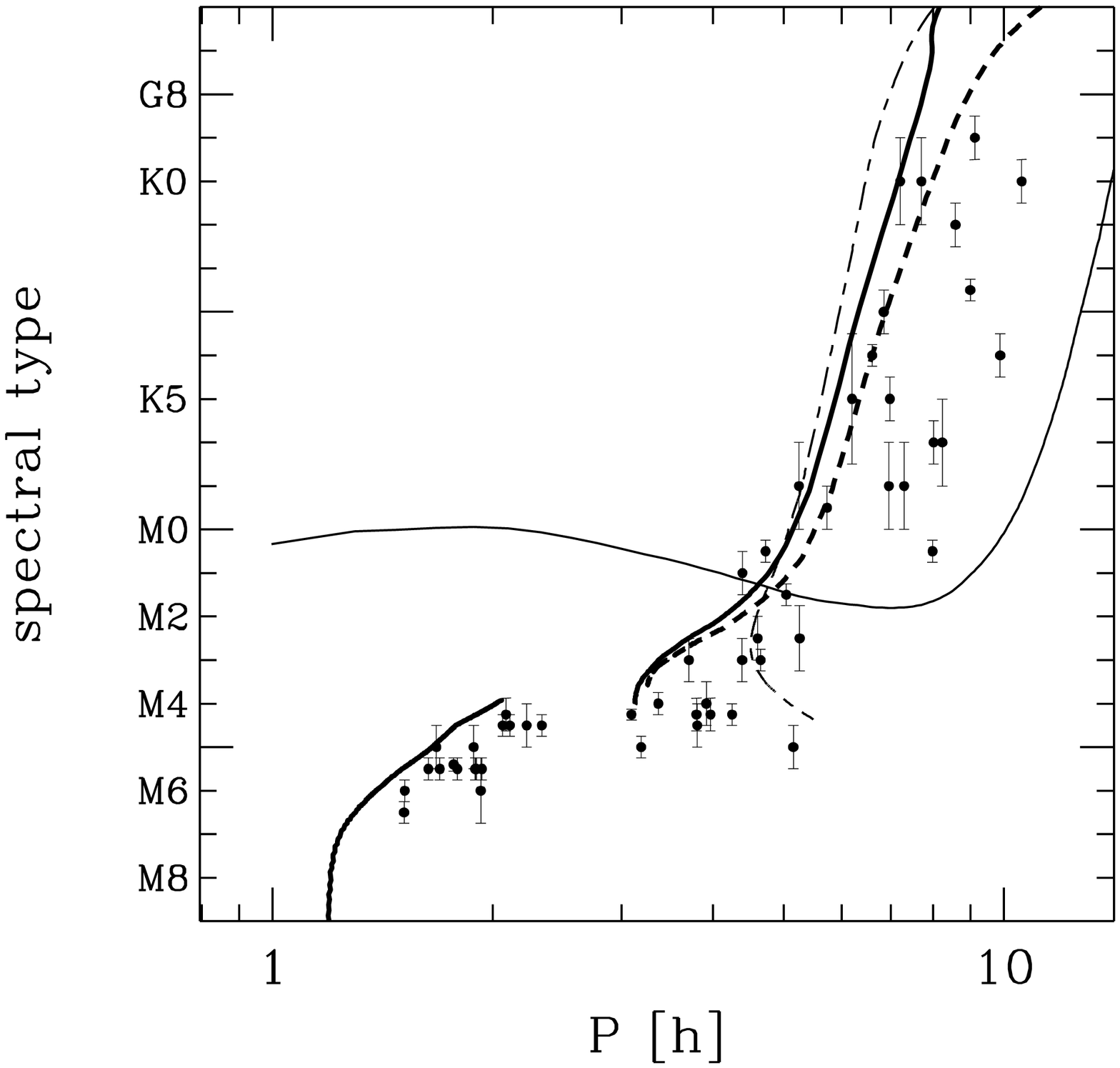,width=7.9cm,angle=0}
\vspace*{-0.5cm}
\caption{Same as Fig.\protect{\ref{fig:sppdata}}, but with various
evolutionary sequences, cf.\ also the track morphology in
Fig.~\protect{\ref{fig:sppmorph}}. Thick solid: standard sequence; thin
solid: evolved low $\mdot$ sequence ($X_c=4\times10^{-4}$); short/long
dashes: unevolved sequence ($\mdot = 10^{-8}\msunyr$);
thick dashes: evolved high $\mdot$ sequence ($X_c=4\times10^{-4}$).
\label{fig:sppmodel}
}
\end{minipage}
\end{figure}

A second invaluable diagnostic diagram for CVs is the orbital
period--spectral type diagram (PS diagram) --- the
``HR diagram'' analogue for CV secondary stars.  
In a semi--detached binary the Roche lobe filling star's mean density
$\rho$ determines the orbital period $P$ almost uniquely (e.g.\ King
1988), $P_h=k/\rho_\odot^{1/2}$, with $k \simeq 8.85$ being only a weak
function of the mass ratio, $P_h$ the period in hr, and $\rho_\odot$
the mean density in solar units. Hence the PS diagram probes 
the donor structure in terms of two stellar parameters that are
relatively easy to determine from observation. 
(The importance of the PS diagram
has already been pointed out by Ritter 1994).   

The ZAMS track defines an upper bound for the scatter of observed CV
secondaries in Fig.~\ref{fig:sppdata}. Systems immediately above and
below the period gap have the same spectral type. 
The results above suggest that these CVs have the same donor mass,
consistent with the standard period gap model according to which CVs
evolve through the gap with constant secondary mass.  

The location of various evolutionary tracks above the period gap 
are shown in Fig.~\ref{fig:sppmorph}. Most tracks intersect the ZAMS
track at a critical period $\pcrit \simeq 5 - 6$~hr, which corresponds
to a donor with mass $M_2\simeq0.6$ $\msun$ marking the transition
from mainly radiative to predominantly convective stars. 
A simple corollary from the results of Sec.~2 is that the donor mass 
on any evolutionary track is about the same as the one on the ZAMS
track with the same spectral type. Deviations to shorter or longer
periods from the ZAMS track mean that the donor is smaller or 
larger then the corresponding ZAMS star, respectively.
From the morphology of tracks in Fig.~\ref{fig:sppmorph} it is clear
that unevolved high $\mdot$ sequences are excluded at long periods
($P\ga\pcrit$), 
but would explain the observed scatter for $P\la\pcrit$. Likewise,
the observations exclude evolved low $\mdot$ sequences at short
periods, while these are clearly required for $P\ga\pcrit$. 

This suggests that, overall, a secular mean mass transfer
rate increasing with decreasing orbital period would provide a good
fit to the observed PS diagram above the period gap. CVs would evolve
along a low $\mdot$ track at long periods and along a high $\mdot$
track at short periods.  
To explain the scatter
of systems at long periods the full spectrum of nuclear evolution
prior to mass loss is required. Note that the sequence starting mass
transfer from a star at the terminal main--sequence defines a lower
bound for the data points in Fig.~\ref{fig:sppmodel}. It is difficult
to see why CVs with somewhat evolved donors should dominate the
population.  
One possibility is that unevolved systems are more susceptible to
irradiation--driven mass transfer cycles (King et al.\ 1996), with
a high state $\mdot$ too high to be recognisable as CVs.
The scatter at short periods ($3-5$~hr) in turn seems to imply a
wide range of secular mean $\mdot$ values, ($10^{-9} - 10^{-8}$
$\msunyr$). None of the orbital angular momentum loss mechanisms
postulated so far generates this wide a 
range of $\mdot$ at a given $P$, while observational estimates for
$\mdot$ always suggest that there is an even larger range (e.g.\
Patterson 1984, Warner 1995).

\centerline{
\psfig{figure=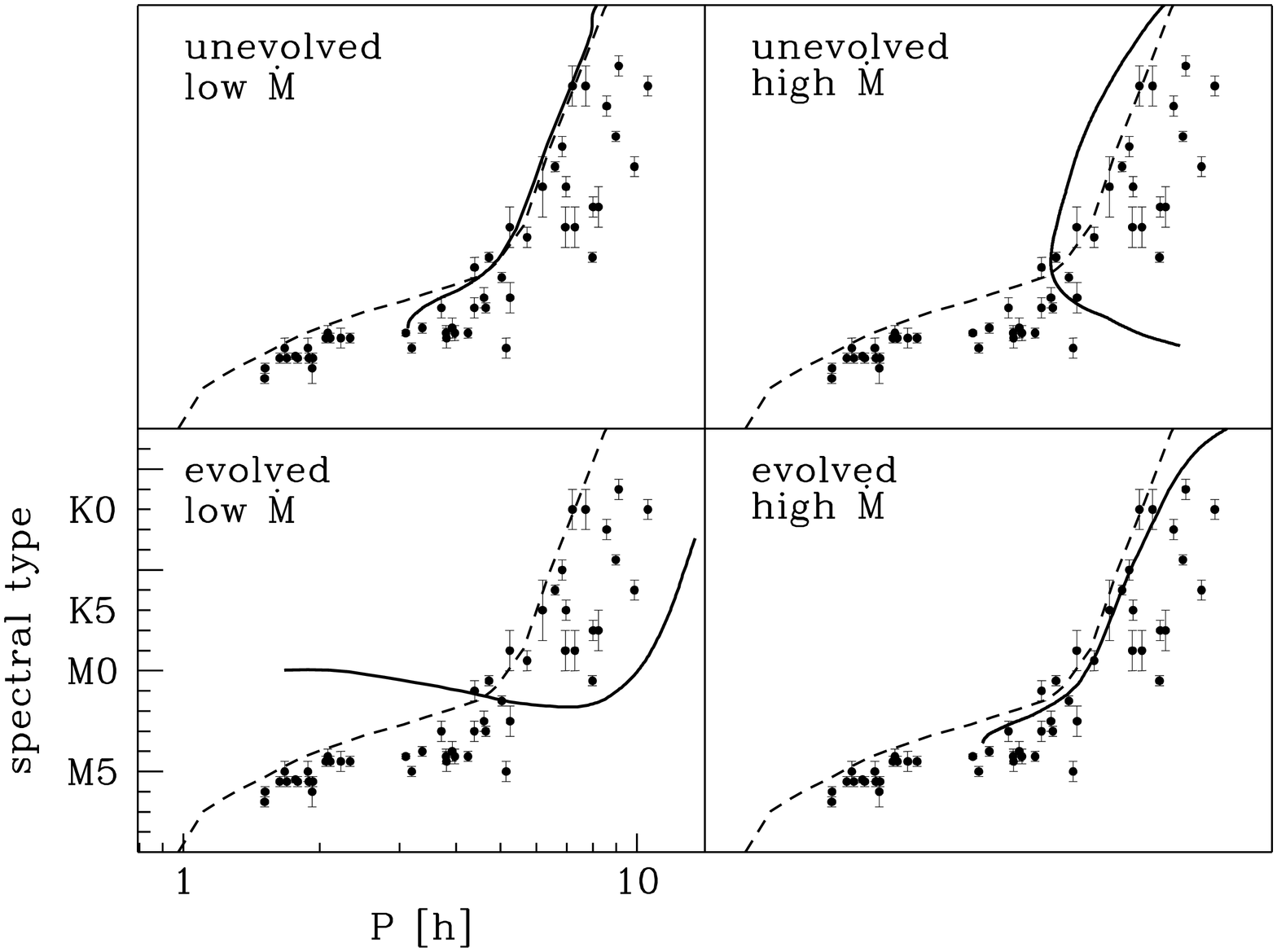,width=16cm,angle=-90}
}

\vspace*{-1.2cm}

\begin{figure}[ht]
\caption{Morphology of evolutionary tracks in the spectral
type--orbital period diagram. Dashed: location of ZAMS
secondaries. Solid curves: standard sequence (upper left panel),
unevolved sequence with $\mdot=10^{-7}$ $\msunyr$ (upper right),
evolved sequence ($X_c=4\times10^{-4}$) with $\mdot = 1.5 \times
10^{-9}$ $\msunyr$ (lower left), and with $\mdot = 5 \times 10^{-8}$ 
$\msunyr$ (lower right).  
\label{fig:sppmorph}
}
\end{figure}

If the $\mdot$ range reflects the range of the secular mean, then the
appearance of a gap with relatively sharp edges is more difficult to
understand than with a uniform evolution driven by
angular momentum losses insensitive to system parameters (Kolb 1996;
Stehle et al.\ 1996).
Yet the observed period distribution and its explanation by disrupted
orbital braking cannot unambiguously rule out a large range of the
secular mean transfer rate above the gap (Baraffe \& Kolb 1999). 
The $\mdot$ range certainly has to level off rather steeply below
$10^{-9}$ $\msunyr$ to avoid filling the gap with systems that hardly
detach. On the other hand, systems with $\mdot \ga 2\times 10^{-9}$
$\msunyr$ undergo period bounce before they reach the upper edge of the
gap, and before the donor becomes fully convective (at a mass
$\mconv$). This bounce period increases with $\mdot$, while $\mconv$
decreases. Hence the detached phase of high $\mdot$ systems extends
from periods longer than 3 hr to periods shorter than 2 hr, and does
not interfere with the ``classical'' gap. The time spent in the 
detached phase increases as well, in the case $\mdot = 10^{-8}$
$\msunyr$ it is longer than 8~Gyr (assuming angular momentum losses by
gravitational radiation). Only a small fraction of such systems 
would reach the orbital period minimum at 80 min within the
age of the Galaxy, thus alleviating (but not solving) the problem of
the missing period spike (see Kolb \& Baraffe 1999) at this minimum
period.

\smallskip

Our conclusions from the mass--spectral type diagram and orbital
period--spectral type diagram can be tested by a set of
more accurate mass and radius determinations of CV secondaries.





\begin{thebibliography}{Davidson \& Humphreys, 1997}
\bibitem[]{}
Allard, F. 1999, in Very Low-Mass Stars and Brown Dwarfs
              in Stellar Clusters and Associations, Euroconference,
La Palma 1998
\bibitem[]{}
Allard, F., Hauschildt, P.H., Baraffe, I., Chabrier, G. 1996,
APJ, 465, L123
\bibitem[]{}
Allard F., Hauschildt P.H., Alexander D.R., Starrfield
S. 1997, ARA\&A, 35, 137
\bibitem[]{}
Baraffe I., Chabrier G. 1996, ApJ, 461, L51
\bibitem[]{}
Baraffe I., Kolb U. 1999, MNRAS, submitted
\bibitem[]{}
Baraffe I., Chabrier G., Allard F., Hauschildt P.H. 1995, ApJ, 446, L35
\bibitem[]{}
Baraffe I., Chabrier G., Allard F., Hauschildt P.H. 1997, A\&A, 327, 1054 
\bibitem[]{}
Baraffe I., Chabrier G., Allard F., Hauschildt P.H. 1998, A\&A, 337, 403
\bibitem[]{} 
Beuermann, K., Baraffe, I., Kolb, U., \& Weichhold, M. 1998,
A\&A, 339, 518
\bibitem[]{}
Chabrier G., Baraffe I. 1995, ApJ, 451, L29
\bibitem[]{}
Chabrier G., Baraffe I. 1997, A\&A, 327, 1039
\bibitem[]{}
Kalogera V., Kolb U., King A.R. 1998, ApJ, 504, 967
\bibitem[]{}
King A.R. 1988, QJRAS, 29, 1 
\bibitem[]{}
King A.R., Kolb U. 1997, ApJ, 481, 918
\bibitem[]{}
King A.R., Frank J., Kolb U., Ritter H. 1996, ApJ, 467, 761
\bibitem[]{}
Kolb U. 1996, in Cataclysmic Variables and
Related Objects, ed.\ A.~Evans, J.H.~Wood, Dordrecht:
Kluwer, IAU Coll.~158, 433
\bibitem[]{}
Kolb U., Baraffe I. 1999, MNRAS, submitted
\bibitem[]{}
Kolb U., Ritter H. 1992, A\&A, 254, 213 
\bibitem[]{}
Kolb U., King A.R., Baraffe I., 1999, MNRAS, submitted
\bibitem[]{}
de Kool M. 1992, A\&A, 261, 188
\bibitem[]{}
Mazzitelli I. 1989, ApJ, 340, 249
\bibitem[]{}
Patterson J. 1984, ApJS, 54, 443
\bibitem[]{}
Politano M. 1996, ApJ, 465, 338
\bibitem[]{}
Ritter H., 1994, in Evolutionary Links in the Zoo of
Interactive Binaries, ed.\ F.~D'Antona, V.~Caloi, C.~Maceroni, 
F.~Giovanelli, in Memorie della Societ\`a Astronomica Italiana 65, 173
\bibitem[]{}
Saumon, D., Chabrier, G., and VanHorn, H.M., 1995, ApJS, 99, 713
\bibitem[]{}
Saumon, D., Chabrier, G., Wagner, D.J., Xie, X., 1998, Phys.\ Rev.,
submitted 
\bibitem[]{}
Smith D.A., Dhillon V.S. 1998, MNRAS, 301, 767
\bibitem[]{}
Stehle R., Ritter H., Kolb U. 1996, MNRAS, 279, 581  
\bibitem[]{}
Warner B. 1995, Cataclysmic Variable Stars, Cambridge Astrophysics
Series 28, Cambridge: CUP
\bibitem[]{}
Whyte C., Eggleton P.P. 1980, MNRAS, 190, 801
\end{thebibliography}
\end{document}